\def\XXint#1#2#3{{\setbox0=\hbox{$#1{#2#3}{\int}$}
\vcenter{\hbox{$#2#3$}}\kern-.5\wd0}}

\documentclass[aps,pre,twocolumn,superscriptaddress,showpacs,preprintnumbers,amsmath,amssymb]{revtex4}

\usepackage{graphicx}
\usepackage{dcolumn}
\usepackage{bm}
\usepackage{color}

\begin{document}

\preprint{}

\title{Generalized Green-Kubo formula and fluctuation theorem for quantum current system around a nonequilibrium steady state}

\author{Hisao Hayakawa}
\affiliation{%
Yukawa Institute for Theoretical Physics, Kyoto University, Sakyo-ku, Kyoto, Japan
}%

\date{\today}

\begin{abstract}
Fluctuation theorem is derived for a quantum current system around a nonequilibrium steady state. 
It is demonstrated that the fluctuation theorem can be a part of the generalized Green-Kubo formula or a nonlinear response theory of an external field or a change of the chemical potential difference. 

\end{abstract}




\maketitle


Green-Kubo formula\cite{kubo} is one of the most fundamental relationships in nonequilibrium statistical mechanics.
The original derivation is restricted to the linearly nonequilibrium case, 
but there exist various generalizations.\cite{evans,zubarev,saito07,fujii,shimizu}
Recently, we have recognized that the fluctuation theorem is a nonlinear generalization of Green-Kubo formula,
and is closely related to some identities such as Jarzynski equality.\cite{FT-papers,Jarzynski97,crooks,Evans02,Seifert05,FT-papers-2,haenggi}
The fluctuation theorem can be applied to various fields and is important 
in particular for mesoscopic physics.

So far, the fluctuation theorem may be regarded as a nonlinear response theory around an equilibrium state. 
Indeed, one of the key steps is to derive the fluctuation theorem 
is to assume a local detailed balance condition.
Thus, nonlinear response theory around a nonequilibrium steady state is outside the existing fluctuation theorem.

The purpose of this paper is 
to generalize the fluctuation theorem around a nonequilibruim steady state.
The formulation is a quantum extension of works by Chong et al.\cite{chong} and  by Evans and Morriss\cite{evans} as well as a preliminary formulation by the present author.\cite{PTPS184-543}



Let us consider a system coupled with two heat baths in which one of the chemical potentials is higher than
another. 
The total Hamilitonian $H_{\rm tot}(t)$ consists of three parts; the system Hamiltonian $H_{\rm S}(t)$, the bath Hamiltonian $H_{\rm B}$ and the interaction Hamiltonian $H_{\rm int}(t)$ as
\begin{equation}\label{5.1}
H_{\rm tot}(t)=H_{\rm S}(t)+H_{\rm B}+g H_{\rm int}(t),
\end{equation}
where $g$ is the coupling constant.

The setup we consider in this paper is as follows: At first, we prepare an equilibrium system as the initial condition in which there is no coupling between the system and the bath, i.e. $H_{\rm int}(t)=0$ at $t\to -\infty$.
Then, we wait for the system to reach a nonequilibrium steady state after an adiabatically connecting between the system and the bath  until $t=0$.
At $t=0$, we may disturb the system and investigate a response to this disturbance.

For later discussion, let us consider a nonlinear response around a nonequilibrium steady state obtained from a coupling between the system and the bath.
Equation (\ref{5.1}) is the most general form of the Hamiltoninan, 
but it is difficult to extract some useful information based on this treatment.
Indeed,  
 a density matrix for a nonequilibrium steady state is necessary for the description of heat conduction and the electric conduction under
a chemical potential difference, but it seems to be impossible to obtain the steady density matrix for Eq.(\ref{5.1}) .
It is remarkable, however,  that Fujii\cite{fujii} derived a steady McLennan-Zubarev form\cite{zubarev,McLennan} of the density matrix 
under a special setup within the framework of Keldysh formalism.\cite{kadanoff-baym,keldysh,kita}

Here, we outline the derivation of the McLennan-Zubarev density matrix within the framework of Keldysh formalism.
Here, let us consider the case that  the reservoirs  is adiabatically connected with the system as
\begin{equation}
H_{\rm tot}^{\epsilon}(t)=H_0+g e^{-\epsilon |t|}H_{\rm int} ,
\label{connection}
\end{equation}
where $H_0\equiv H_{\rm S}+H_{\rm B}$ under no time dependence in $H_0(t)$ and $H_{\rm int}(t)$,  
$H^{\epsilon}_{\rm tot}(t)\to H_{\rm tot}(t)$ in the limit $\epsilon\to 0$, and 
the coupling between the system $H(t)$ and the bath $H_{\rm B}$ is adiabatically  zero in $|t|\to \infty$.
Note that we always take the limit $\epsilon\to 0$ for later discussion, even when we do not write its limit explicitly.
 
Here, the density matrix at $t=0$ is given by
\begin{equation}\label{S_matrix}
\bar{\rho}^{\epsilon}=S^{\epsilon}(0,-\infty)\rho^0S^{\epsilon}(-\infty,0) ,
\end{equation}
where $S^\epsilon(t,t_0)\equiv T_{\leftarrow}\exp[-\frac{i}{\hbar} g \int_{t_0}^tdse^{-\epsilon |s|}H_{\rm int}(s)]$ for $t>t_0$ and
$S^\epsilon(t,t_0)\equiv T_{\rightarrow}\exp[-\frac{i}{\hbar} g \int_{t_0}^tdse^{-\epsilon |s|}H_{\rm int}(s)]$ for $t<t_0$
with the introduction of the right ordered operator $T_\rightarrow$ and the left ordered operator $T_\leftarrow$.
Here, the initial condition of the total system in the limit $t\to -\infty$ is assumed to be
\begin{equation}\label{initial-total}
\rho^0=e^{-\beta (H_0-e\Phi  \Delta N/2)}/Z_{\rm tot} ,
\end{equation}
where  $e$ is the charge of an electron, $\Phi$ is the voltage difference between two reservoirs,
and $\Delta N$ is the number difference of electrons between two reservoirs. 
To reproduce this initial density matrix (\ref{initial-total}) we have assumed that the average 
$(\mu_{\rm L}+\mu_{\rm R})/2$ 
of the left chemical potential $\mu_{\rm L}$ and the right chemical potential $\mu_{\rm R}$ 
is zero, and $\Phi$ satisfies $e\Phi=\mu_{\rm L}-\mu_{\rm R}$.
Note that the expression (\ref{S_matrix}) is only valid for the case of $[H_0,\rho^0]=0$.

As was shown in Ref.\cite{fujii}, it is notable that any observable $A$ commutable with $H_0$ at $t=0$ satisfies
\begin{eqnarray}\label{fujii42}
\bar{A}^{\epsilon}&=& S^{\epsilon}(0,-\infty)AS^{\epsilon}(-\infty,0)=A+\int_{-\infty}^0dt e^{-\epsilon|t|} J^A_H(t) \nonumber\\
&=&\epsilon \int_{-\infty}^0dte^{-\epsilon |t|}A_H(t) ,
\end{eqnarray}
where we have introduced $J_H^A(t)\equiv -(\partial/\partial t) A_H(t)$.  
Note that $\bar{A}^{\epsilon}$ in Eq. (\ref{fujii42}) is equivalent to the invariant measure introduced by Zubarev.\cite{zubarev}

It is straightforward to obtain the density matrix at $t=0$ as\cite{fujii}
\begin{eqnarray}
\bar{\rho}^{\epsilon}(0)
&=& \exp\left\{-\beta\left[H_0^{\epsilon}(0)-\frac{e\Phi}{2}\Delta N^{\epsilon}(0)\right]\right\}/Z_{\rm tot}
\label{fujii66}
\end{eqnarray}
in the limit $\epsilon \to 0$, where 
\begin{eqnarray}
H_0^\epsilon(0)&\equiv& H_0+\int_{-\infty}^0dte^{-\epsilon |t|}{J}_{e,H}(t) , \nonumber\\
\Delta N^{\epsilon}(0) &\equiv&
\Delta N+\int_{-\infty}^0dt e^{-\epsilon |t|}J_{c,H}(t)
\end{eqnarray}
 with the energy current
$J_{e,H}^{\epsilon}(t)\equiv -(\partial/\partial t) H_{0,H}(t)$ 
and the mass current $J_{c,H}^\epsilon(t)\equiv -(\partial/\partial t) \Delta N_H(t)$.
The density matrix (\ref{fujii66}) is equivalent to the classical nonequilibrium density matrix by McLennan-Zubarev\cite{zubarev,McLennan}.
For later convenience, we rewrite Eq.(\ref{fujii66}) as
\begin{equation}\label{steady_total}
\bar{\rho}^{\epsilon}(0)
=\frac{\exp\left\{-\beta {\cal H}^\epsilon(0) \right\}}{Z^\epsilon(\beta)},
\end{equation}
where ${\cal H}^\epsilon(0) \equiv H_0^\epsilon(0)-e(\Phi/2)\Delta N^\epsilon(0)$
and $Z^\epsilon(\beta)={\rm tr}e^{-\beta{\cal H}^{\epsilon}(0)}$. 
This is the consequence of Keldysh Green function formalism for a nonequilibrium steady state.\cite{fujii}
It is straightforward to introduce $\bar{\rho}^\epsilon(t)$ at $t$ satisfying $|t|<1/\epsilon$ as the generalization of Eq.(\ref{fujii66}).
It should be noted that the density matrices $\bar{\rho}_{\rm tot}^\epsilon(t)$ is in a steady state in time $|t|\ll 1/\epsilon$.

We should note that von-Neumann equation for the total system 
\begin{equation}\label{5.2}
\frac{d}{dt}\bar{\rho}^\epsilon(t)=-i {\cal L}_{{\rm tot},0}^{\epsilon}(t) \bar{\rho}^\epsilon(t) \equiv 
\frac{1}{i\hbar}[H_{\rm tot}^{\epsilon}(t),\bar{\rho}^\epsilon(t)]
\end{equation}
for the  density matrix $\bar{\rho}_{\rm tot}^\epsilon(t)$ 
satisfies
\begin{equation}
\frac{d\bar{\rho}^{\epsilon}(t)}{dt}=i{\cal L}_{{\rm tot},0}^\epsilon(t) \bar{\rho}^\epsilon(t)=0 
\end{equation}
for $|t|\ll 1/\epsilon$, where the subscript 0 for the Liouville operator ${\cal L}_{0}^{\epsilon}(t)$ is introduced to represent a reference state for the response theory for later discussion.
Because of the relation $[\bar{\rho}^\epsilon(0),H_0^\epsilon(0)]=[\bar{\rho}^\epsilon(0),\Delta N^\epsilon(0)]=0$ the  density matrix $\bar{\rho}^\epsilon(0)$ 
also satisfies 
\begin{equation}\label{steady_eq}
i{\cal L}^{\epsilon}_{0}(t) \bar{\rho}^\epsilon(t)=0 
\end{equation}
for $|t|\ll 1/\epsilon$, 
where the Liouville operator ${\cal L}^{\epsilon}_{0}(t)$  can be formally written as
\begin{equation}\label{L_dagger}
i {\cal L}^{\epsilon}_0(t)=\frac{i}{\hbar}[{\cal H}^\epsilon(t),\quad ] 
\end{equation}
for $|t|\ll 1/\epsilon$. 
It should be noted that the Liouville operator ${\cal L}_0^\epsilon(t)$ is unitary. 


Now, let us discuss how we can construct a nonlinear response theory around the nonequilibrium steady state Eq.(\ref{steady_total}).
To avoid heavy notations, we drop the adiabatic factor $\epsilon$ of any variables for later discussion under the limit $\epsilon \to 0$. Therefore, though our argument is only valid $0<t<1/\epsilon$, the formulation can be used for any finite $t$ in the limit $\epsilon \to 0$.
We should note that $i {\cal L}_0(t)=\lim_{\epsilon \to 0}i{\cal L}_0^{\epsilon}(0)$ for $|t|\ll 1/\epsilon$.
We consider response theories after the following two cases  for $t\ge 0$:
\begin{eqnarray} \label{caseA}
\Phi(t)&=& \Phi+ \Delta \Phi(t), \qquad (\rm{ case{~} A} ) \\
H_{0}(t)&=&H_{0}(0)+ F_{\rm ex}(t) B \qquad (\rm{ case{~} B} )
\label{caseB}
\end{eqnarray} 
where $B$ is the conjugate field to the external force $F_{\rm ex}(t)$.

Let us introduce the difference of the Liouville operators $i\Delta {\cal L}(t)\equiv i{\cal L}(t)-i{\cal L}_0(0)$ 
 from the base state as
\begin{equation}\label{diff_A}
i \Delta{\cal L}^{({\rm A})}(t)=-\frac{i e \Delta \Phi(t)}{2\hbar}\left[\Delta N,\quad \right]
\end{equation}
for case A and
\begin{equation}\label{diff_B}
i \Delta {\cal L}^{({\rm B})}(t)=\frac{i F_{\rm ex}(t)}{\hbar}[B, \quad]
\end{equation}
for case B, where $i{\cal L}(t)$ is the Liouville operator for $t\ge 0$, and
the upperscripts (A) and (B) are introduced corresponding to the cases A and B, respectively.
(To represent both cases we do not put any upperscript).
By using $i\Delta {\cal L}(t)$, the time evolution of the density matrix of the system at $t\ge 0$ can be described by
\begin{equation}\label{time-evolve-density}
\frac{d}{dt}\bar{\rho}(t)=-i\Delta {\cal L}(t) \bar{\rho}(t) ,
\end{equation}
and its formal solution is given by
\begin{equation}\label{density_matrix}
\bar{\rho}(t)=U_{\leftarrow}(t,0)\bar{\rho}(0)
\end{equation}
where 
\begin{equation}\label{def_tilde_U}
{U}_{\leftarrow}(t,0)=T_{\leftarrow}[e^{-i\int_{0}^tds\Delta {\cal L}(s)}].
\end{equation}
These results are based on Schr\"{o}dinger picture.

We can also introduce the adjoint dynamics for observable $A_H(t)$ based on the Heisenberg picture as
\begin{equation}\label{heisenberg}
\frac{d}{dt}A_H(t)
=U_{\rightarrow}(-\infty,t)i {\cal L}(t) A,
\end{equation}
where the subscript $H$ represents an observable in Heisenberg picture, and we have introduced
\begin{equation}\label{U_right}
U_\rightarrow(t_0,t)\equiv
T_{\rightarrow} e^{i\int_{t_0}^t ds {\cal L}(s)}.
\end{equation}
It is possible to prove the relation\cite{evans}
\begin{equation}
U_\rightarrow(t_1,t_2)=U_\rightarrow(t_1,\tau)U_\rightarrow(\tau,t_2) 
\end{equation}
and
\begin{equation}
U_\rightarrow(t_0,t)=U_\leftarrow(t,t_0)^{-1} .
\end{equation}
The formal solution of Eq.(\ref{heisenberg}) is given by
\begin{equation}
A_H(t)=U_\rightarrow(0,t)A_H(0)= U_\rightarrow(0,t)A_H(0)U_\leftarrow(t,0),
\end{equation}
where $A_H(0)=U_{\rightarrow}^0(-\infty,0)A=T_{\rightarrow} e^{i\int_{-\infty}^0ds {\cal L}_0(s)}A$.
The last expression can be proved with the aid of Baker-Campbell-Hausdorff formula. 
Here, $U_\rightarrow(0,t)$ in Eq. (\ref{U_right}) can be rewritten as
\begin{equation}
U_\rightarrow(0,t)=T_\rightarrow e^{\int_0^tds i\Delta{\cal L}(s)} ,
\end{equation}
because $A_H(0)$ also satisfies the steady condition
\begin{equation}
\label{steady_cond}
U_\rightarrow^0(0,t)A(0)=A(0)
\end{equation}
with $A(0)\equiv A_H(0)$ and 
$U_\rightarrow^0(0,t)=T_\rightarrow e^{\int_0^tds i{\cal L}_0(s)}$.
Note that there are two relations
\begin{equation}\label{unchange}
B_H(t)=B, \quad {\rm and} \quad \Delta N_H(t)=\Delta N.
\end{equation}

Therefore, the expectation value of the observable $A$ based on the steady density matrix $\bar{\rho}(0)$ defined in Eq. (\ref{fujii66}) is represented by
\begin{equation}
\langle A_{\rm H}(t)\rangle\equiv {\rm tr}_S\{\bar{\rho}(0)A_H(t)\}
={\rm tr}\left\{\bar{\rho}(0)U_\rightarrow(0,t)A_H(0) \right\}.
\end{equation}
The Dyson equation between $U_\rightarrow(0,t)$ and $U_\rightarrow^0(0,t)$ can be written as
\begin{equation}\label{Dyson_eq2}
{U}_\rightarrow(0,t)
=U_\rightarrow^0(0,t)+\int_0^tds U_\rightarrow^0(0,s)i \Delta {\cal L}(s) U_\rightarrow(s,t) .
\end{equation}
From the steady condition (\ref{steady_cond})
we obtain
\begin{eqnarray}
& &\langle \delta A_H(t)\rangle\nonumber\\
&& \quad
= 
\int_0^t d\tau {\rm tr}
\left\{
 \bar{\rho}(0) 
U_\rightarrow^0(0,\tau)i \Delta {\cal L}(\tau) U_\rightarrow(\tau,t) A_H(0)
\right\}
\nonumber\\
&& \quad = 
\int_0^td\tau {\rm tr}\left\{ \left[U_\leftarrow^0(\tau,0)\bar{\rho}(0) \right]
i \Delta {\cal L}(\tau) U_\rightarrow(\tau,t) A_H(0) 
\right\}
\nonumber\\
&&\quad =
\int_0^t d\tau {\rm tr}
\left\{
\bar{\rho}(0)i \Delta {\cal L}(\tau) U_\rightarrow(\tau,t)A_H(0)
 \right\} ,
\label{<A>}
\end{eqnarray}
where $\langle \delta A_H(t) \rangle\equiv \langle A_H(t) \rangle-\langle A_H(0) \rangle $, and
we have also used the equivalency between Schr\"{o}dinger picture and Heisenberg picture in the second line
and the steady condition $U_\leftarrow^0(\tau,0)\bar{\rho}(0)=\bar{\rho}(0)$ for the final expression.
Equation (\ref{<A>}) can be rewritten as
\begin{equation}
\langle \delta A_H(t) \rangle
=
\int_0^t d\tau {\rm tr}
\left\{
\left[
i \Delta {\cal L}(\tau) \bar{\rho}(0)
\right]
 U_\rightarrow(\tau,t)A_H(0)
 \right\} 
\label{<A>2} 
\end{equation}
in terms of the equivalency of two pictures.

Because of the relation (\ref{fujii66}) and (\ref{unchange}),  
 Eq.(\ref{diff_A}) is
rewritten as
\begin{eqnarray}\label{diff_A2}
i \Delta{\cal L}_{\rm A}(t) \bar{\rho}(0)
&=&-\frac{i e \Delta \Phi(t)}{2\hbar}\left[\Delta N_H(t), \bar{\rho}(0)\right]
\nonumber\\
&=&
-\frac{ e \Delta \Phi(t)}{2}
\bar{\rho}(0) \int_0^{\beta}d\lambda e^{\lambda {\cal H}(0)}
J_H^{(N)}(t)
 e^{-\lambda {\cal H}(0)}
\nonumber\\
&=&
-\frac{ e \Delta \Phi(t)}{2} \bar{\rho}(0)
\int_0^{\beta} d\lambda \breve{J}_H^{(N)} (t;-i\hbar\lambda) ,
\end{eqnarray}
where we have used Kubo's identity in  the second line and we have introduced 
$ J_H^{(N)}(t) \equiv 
1/(i\hbar)
[\Delta N_H(t),{\cal H}]
$
and  $\breve{J}_H^{(N)}(t;\tau)\equiv
e^{-i{\cal H}(0)\tau/\hbar}J_H^{(N)}(t) e^{-i{\cal H}(0) \tau/\hbar}$.
Similarly, Eq.(\ref{diff_B}) can be rewritten as
\begin{equation}\label{diff_B2}
i \Delta{\cal L}_{\rm B}(t) \bar{\rho}(0)
= F_{\rm ex}(t) \bar{\rho}(0) \int_0^{\beta} d\lambda \breve{J}_B(t;-i\hbar\lambda) ,
\end{equation}
where 
$
J_H^{(B)}(t) \equiv 
1/(i\hbar)
[B_H(t),{\cal H}(0)]
$
and $\breve{J}_H^{(B)}(t;\tau)\equiv
e^{-i{\cal H}(0)\tau/\hbar}J_H^{(B)}(t) e^{-i{\cal H}(0) \tau/\hbar}$.

Because case A and case B are almost equivalent with each other because of the comparison between Eqs.(\ref{diff_A2}) and (\ref{diff_B2}),
it is sufficient to prove one of two cases.
In this paper, let us analyze the case B.
Substituting Eqs.(\ref{steady_eq}) and (\ref{diff_B2}) into Eq.(\ref{<A>2})
we obtain
\begin{eqnarray}\label{Feb12_21}
 \langle \delta A_H(t) \rangle
&=& -\int_0^td\tau F_{\rm ex}(\tau)\int_0^\beta d\lambda
\nonumber\\
&& \quad
\langle
\breve{J}_H^{(B)}(\tau;-i\hbar\lambda)U_\rightarrow(\tau,t)A_H(0)
\rangle.
\end{eqnarray}
With the aid of $U_\rightarrow(\tau,t)=U_\rightarrow(0,\tau)^{-1}U_\rightarrow(0,t)=U_\leftarrow(\tau,0)U_\rightarrow(0,t)$
Eq.(\ref{Feb12_21}) is further rewritten as
\begin{eqnarray}\label{Feb12_25}
\langle \delta A_H(t)\rangle
&=&
-\int_0^td\tau F_{\rm ex}(\tau)
\langle \breve{J}_H^{(B)}(\tau;-i\hbar\lambda)U_\leftarrow(s,0)A_H(t) \rangle
\nonumber\\
&=&
-\int_0^td\tau F_{\rm ex}(\tau) \int_0^\beta d\lambda
{\rm tr}\LARGE\{U_\rightarrow(0,\tau)
\nonumber\\
&& \qquad
\left[
\bar{\rho}(0)
\breve{J}_H^{(B)}(\tau;-i\hbar\lambda)
\right]
A_H(t) \LARGE\} .
\end{eqnarray}
With the aid of Eq.(\ref{unchange}) we have the invariant relation $J_H^{(B)}(t)=J_H^{(B)}$. Then,
we may have the relation
\begin{eqnarray}
&& U_\rightarrow(0,t)[\bar{\rho}(0)\breve{J}_H^{(B)}(t;-i\hbar\lambda)]
\nonumber\\
&&\quad =
U_\rightarrow(0,t)\bar{\rho}(0)\cdot U_\rightarrow(0,t)\breve{J}_H^{(B)}(t;-i\hbar\lambda)
\nonumber\\
&& \quad =
U_\rightarrow(0,t)\bar{\rho}(0)\cdot \breve{J}_H^{(B)}(t;-i\hbar\lambda) .
\label{keisan}
\end{eqnarray}
Here,
from Eqs.(\ref{def_tilde_U}) and (\ref{diff_B2}) we can write
\begin{equation}
U_\rightarrow(0,t)\bar{\rho}(0)=\bar{\rho}(0) T_\rightarrow e^{-\int_0^td\tau \Omega(\tau) } ,
%
\label{tilde_U_2}
\end{equation} 
where $\Omega(\tau) \equiv -F_{\rm ex}(\tau)\int_0^{\beta} d\lambda \breve{J}_H^{(B)}(\tau;-i\hbar\lambda)$.
Substituting Eq.(\ref{tilde_U_2}) into (\ref{Feb12_25}) we obtain
\begin{equation}\label{g-Green-Kubo}
 \langle \delta A_H(t) \rangle
=\int_0^tds F_{\rm ex}(s) \int_0^{\beta} d\lambda 
\Phi_{BA}(s)
,
\end{equation}
where $\Phi_{BA}(t)\equiv \langle T_{\rightarrow} e^{-\int_0^t d\tau \Omega(\tau)}
\breve{J}_H^{(B)}(t;-i\hbar\lambda)A_H(t) \rangle$ is the nonlinear response function.
Equation (\ref{g-Green-Kubo}) is the quantum version of the generalized Green-Kubo formula.\cite{evans}
If we replace the observable $A$ by the current $J$, the expression (\ref{g-Green-Kubo}) can be rewritten as
\begin{equation}\label{g-GK-2}
\langle \delta J_H(t) \rangle
=\int_0^tds F_{\rm ex}(s) \int_0^{\beta} d\lambda 
 \Phi_{BJ}((s),
\end{equation}
where we have introduced $\langle \delta J_H(t) \rangle\equiv \langle J_H(t) \rangle - \langle J(0) \rangle$.
These expressions are the quantum version of the nonlinear response theory obtained by Evans and Morriss.\cite{evans}
Needless to say, it is reduced to the conventional linear response formula\cite{kubo} if we set $\Omega(\tau)=1$.
It should be noted that the nonlinear response theory in case A can be obtained when we replace $\breve{J}_H^{(B)}(\tau;-i\hbar\lambda)$ by $\breve{J}_H^{(N)}(\tau;-i\hbar \lambda)$.

The fluctuation theorem is easily obtained from the normalization condition ${\rm tr}\bar{\rho}(-t)=1$ as
\begin{equation}\label{FT}
\langle
T_\rightarrow e^{-\int_0^td\tau 
\Omega(\tau)}
\rangle
=1 ,
\end{equation}
where we have used $\bar{\rho}(-t)=U_{\rightarrow}(0,t)\bar{\rho}(0)$.
Equation (\ref{FT}) is the integral fluctuation theorem in Refs.\cite{Seifert05,chong}. 
From Jensen's inequality, (\ref{FT}) leads to $\int_0^td\tau\langle \Omega(\tau) \rangle \ge 0$, which means that
“entropy production" is positive through the process. 

It should be noted that our nonlinear response theory of fluctuation theorem has the same form as that around an equilibrium state thanks to the special choice of the nonequilibrium steady state Eq.(\ref{fujii66}).
Of course, we cannot obtain the explicit form of any expectation value within the framework presented here,
 because the steady density matrix Eq.(\ref{fujii66}) contains the history of the current.
 The explicit calculation to demonstrate the usefulness of this framework will be reported elsewhere.

In conclusion, 
we have derived a quantum nonlinear response theory (\ref{g-Green-Kubo}) around a nonequilibrium steady state (\ref{fujii66}) for a dissipative quantum system. 
We have also derived the quantum fluctuation theorem (\ref{FT}) for dissipative quantum systems.
 This formulation is directly applicable to the description of the dynamics of a subsystem charcterized by the system Hamiltoninan $H_{\rm S}$
if the coupling between the system and the bath is sufficiently weak. 
Actually, Evans and Morriss\cite{evans} derived the generalized Green-Kubo formula for classical dissipative system under the assist
of the adiabatic incompressibility of the phase space. 
The formulation
should be extended to the case for dissipative subsystems  with finite coupling between the subsystem and the bath.
Although the results presented here is rather formal, the explicit calculation following this formulation will be reported elsewhere. 

{\bf Acknowledgment} The author thanks M.Otsuki, S-H. Chong,  C. Uchiyama, Y. Utsumi and K. Saito for fruitful discussion.
This work is partially supported by the
Global COE Program “The Next Generation of Physics,
Spun from Universality $\&$ Emergence" from the Ministry
of Education, Culture, Sports, Science and Technology
(MEXT) of Japan, the Japan Society for the Promotion
of Science for Young Scientists (JSPS), and the Grant-in-Aid of MEXT (Grants No. 21540384).

\end{document}